\documentclass[sigconf,natbib=true]{acmart}
\AtBeginDocument{%
  \providecommand\BibTeX{{%
    \normalfont B\kern-0.5em{\scshape i\kern-0.25em b}\kern-0.8em\TeX}}}

\copyrightyear{2024}
\acmYear{2024}
\setcopyright{acmlicensed}\acmConference[SIGIR '24]{Proceedings of the 47th International ACM SIGIR Conference on Research and Development in Information Retrieval}{July 14--18, 2024}{Washington, DC, USA}
\acmBooktitle{Proceedings of the 47th International ACM SIGIR Conference on Research and Development in Information Retrieval (SIGIR '24), July 14--18, 2024, Washington, DC, USA}
\acmDOI{10.1145/3626772.3657971}
\acmISBN{979-8-4007-0431-4/24/07}

\DeclareMathOperator{\vect}{vec}
\usepackage{enumitem}
\usepackage{array}
\usepackage{makecell}
\usepackage{multirow}
\usepackage[normalem]{ulem}
\usepackage{balance} 
\useunder{\uline}{\ul}{}

\begin{document}

\title{Masked Graph Transformer for 
Large-Scale Recommendation}

\author{Huiyuan Chen}
\email{hchen@visa.com}
\affiliation{%
  \institution{Visa Research}
  \city{Foster City}
  \country{USA}}

\author{Zhe Xu}
\email{zhexu3@illinois.edu}
\affiliation{%
  \institution{
University of Illinois Urbana-Champaign}
  \city{Illinois}
  \country{USA}}

\author{Chin-Chia Michael Yeh}
\email{miyeh@visa.com}
\affiliation{%
  \institution{Visa Research}
  \city{Foster City}
  \country{USA}}

\author{Vivian Lai}
\email{viv.lai@visa.com}
\affiliation{%
  \institution{Visa Research}
  \city{Foster City}
  \country{USA}}

\author{Yan Zheng \\Minghua Xu}
\email{yazheng@visa.com}
\affiliation{%
  \institution{Visa Research}
  \city{Foster City}
  \country{USA}}

\author{Hanghang Tong}
\email{htong@illinois.edu}
\affiliation{%
  \institution{
University of Illinois Urbana-Champaign}
  \city{Illinois}
  \country{USA}}

\renewcommand{\shortauthors}{Huiyuan Chen et al.}


\begin{abstract}
Graph Transformers have garnered significant attention for learning graph-structured data, thanks to their superb ability to capture long-range dependencies among nodes. However, the quadratic space and time complexity  hinders the scalability of Graph Transformers, particularly for large-scale  recommendation. Here we propose an efficient Masked Graph Transformer, named MGFormer, capable of capturing  all-pair interactions among nodes with a linear complexity. To achieve this, we treat all user/item nodes as independent tokens, enhance them with positional embeddings, and feed them into a kernelized attention module. Additionally, we incorporate learnable relative degree information to appropriately reweigh the attentions. Experimental results show the superior performance of our MGFormer, even with a single attention layer.

\end{abstract}



\maketitle

\section{Introduction}

Graph Neural Networks (GNNs) are widely used in recommender systems, owing to their remarkable performance~\cite{he2020lightgcn,yu2022graph,zhao2024can}. 
Nevertheless, GNNs often encounter difficulties in capturing long-range dependencies because graph convolutions are inherently local operations, limiting their
expressive power~\cite{wu2022nodeformer,huang2023tailoring,chen2022graph}. While deep GNNs can increase the receptive field to learn more complex patterns, training deep GNNs
poses new challenges, such as over-smoothing~\cite{li2018deeper,yan2023from} and high complexity~\cite{wang2022improving,yeh2022embedding,xu2023kernel,zhao2024leveraging}.


Recently, Graph Transformers (GTs) have illuminated the landscape of capturing long-range dependencies,  outperforming GNNs across various graph benchmarks~\cite{rampavsek2022recipe,geisler2023transformers,shirzad2023exphormer,ying2021transformers,ma2023}. The core idea of GTs is the self-attention mechanism so that the message passing operates on a complete attention graph, in contrast to GNNs whose message passing works on the original graph and is hard to capture long-range interactions. In addition, GTs further develop various positional encodings to capture topology information between nodes, including Laplacian positional encodings~\cite{kreuzer2021rethinking}, centrality and spatial encodings~\cite{ying2021transformers}, and random walk structural encodings~\cite{yu2024p,dwivedi2021graph,ma2023}, which can further improve the performance.

However, GTs often suffer from quadratic space and time complexity with respect to the number of nodes~\cite{vaswani2017attention,ying2021transformers,yeh2023toward,lai2023enhancing,jin2024llm}, hindering their applicability for large-scale graphs. In  NLP,  various efforts have been made to enhance the efficiency of Transformers for handling long sequences, including sparse attentions~\cite{zaheer2020big,chen2022denoising}, locality sensitive hashing~\cite{kitaev2019reformer}, and low-rank approximation~\cite{wang2020linformer}. However, it remains unclear how these lightweight Transformers can be effectively applied to large-scale graphs in recommendation, where the graphs may consist of millions of nodes (tokens).

\begin{figure}
\centering
\includegraphics[width=7.0cm]{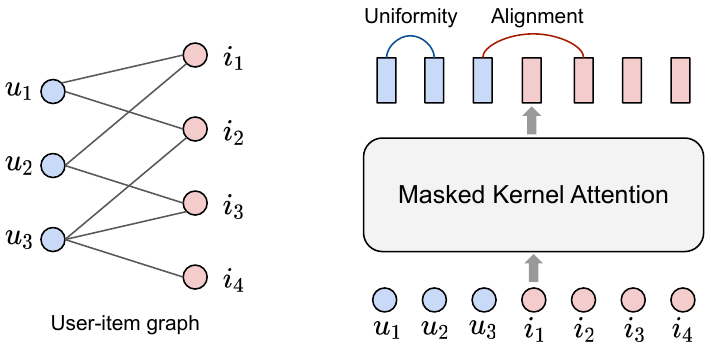}
\caption{Overview of the proposed MGFormer.}
\label{graph}
\vspace{-15pt}
\end{figure}

In this paper, we propose an efficient Masked Graph Transformer, named MGFormer, capable of capturing all-pair interactions among nodes with a linear complexity. Illustrated in Figure~\ref{graph}, we treat each user/item node as an independent token and feed all nodes into our masked kernel attention layer to learn the representations of users and items. In particular, our MGFormer consists of three key ingredients: 1) Structural encoding, leveraging the bipartite graph's topology to discern each node's position, 2) Kernelized attention layer, approximating self attention as a linear dot-product of kernel features with  linear complexity, and 3) Reweighing attention mechanism, utilizing a learnable sinusoidal degree mask to adjust attention distribution, which assigns more credits to more crucial tokens.  The experimental results demonstrate that our MGFormer achieves competitive performance across multiple  benchmarks, while maintaining  a comparable complexity to GNNs.


\section{Related Work}
\textbf{GNNs for Recommendation:} Graph Neural Networks (GNNs) utilize a message-passing mechanism that enables nodes to communicate with their neighbors~\cite{rampavsek2022recipe,geisler2023transformers,chen2021structured,shirzad2023exphormer,song2023,ying2021transformers,ma2023,chen2022structure,chen2023sharpness}. This strong graph inductive bias is crucial for modeling user-item interactions. For example, LightGCN~\cite{he2020lightgcn} learns user and
item embeddings by linearly propagating signals on the graph. UltraGCN~\cite{mao2021ultragcn}  bypasses 
infinite layers of explicit message passing for more efficient  recommendation.  SimGCL~\cite{yu2022graph} adopts a contrastive framework by introducing random noises as data augmentation. Recently, DirectAU~\cite{wang2022towards}  explicitly optimizes alignment and uniformity of user and item representations on the hypersphere.  Despite the remarkable performance of GNNs, their ability to capture long-range dependencies based on local convolution remains insufficient, especially neglecting the impact of low-degree nodes.\\

\noindent\textbf{Graph Transformer:} 
Graph Transformers (GTs) typically calculate full attentions, regardless of the edge connectivity~\cite{rampavsek2022recipe,geisler2023transformers,shirzad2023exphormer,ying2021transformers,ma2023}. This enables  GTs to more effectively capture long-range dependencies compared to GNNs. For example, Graphormer~\cite{ying2021transformers} combines  self-attention mechanism with graph structural encodings. SAT~\cite{chen2022structure} empowers  attention layer with extracting  subgraph representations. Remarkably, both methods demonstrate impressive performance on molecular datasets. In the field of recommendation, GFormer~\cite{Li_2023}  leverages the Transformer to acquire environment-invariant user preference. LightGT~\cite{wei2023lightgt} employs the Transformer to learn modal-specific embeddings for multimedia recommendation.  These studies still incorporate GNN components, thereby potentially inheriting issues such as oversmoothing. In contrast, recent efforts  highlight that a pure Transformer yields a powerful graph learner both in theory and practice~\cite{kim2022pure,wu2022nodeformer,huang2023tailoring}. Here we align with this research direction, aiming to construct a pure and efficient Transformer for large-scale  collaborative filtering. 

\section{METHODOLOGY}
\subsection{Preliminaries}
\subsubsection{\textbf{Problem Setup}}
Following ~\cite{he2020lightgcn,wang2022towards,chen2022structure}, let $\mathbf{R} \in \mathbb{R}^{M \times N}$ be theuser-item interaction matrix, where $M$ and $N$ denote the number of users and items, respectively.  $\mathbf{R}_{ui}$ is set to $1$ when the user $u$ has interacted with the item $i$ previously, and $0$ for unobserved interactions. The goal  is to recommend users a ranked list of items that are unobserved before. 

\subsubsection{\textbf{Self-Attention Mechanism}}

The attention layer is one of the key components in the Transformer~\cite{vaswani2017attention}. Let $\mathbf{X} \in \mathbb{R}^{n \times d}$ denote the input sequence,  where $n$ is the length and $d$ is the dimension size. The input $\mathbf{X}$ is  projected to queries $\mathbf{Q}$, keys  $\mathbf{K}$, and values  $\mathbf{V}$ through linear transformations, $\mathbf{Q} =  \mathbf{X} \mathbf{W}_Q, \mathbf{K} =  \mathbf{X} \mathbf{W}_K$, and  $\mathbf{V} =  \mathbf{X} \mathbf{W}_V$, respectively. The output at each position $\mathbf{h}_i$ is computed as:
\begin{equation}
    \mathbf{h}_i = \frac{\sum_{j=1}^{n}\text{sim}( \mathbf{q}_i, \mathbf{k}_j)  \cdot \mathbf{v}_j}{\sum_{j=1}^{n}\text{sim}( \mathbf{q}_i, \mathbf{k}_j)}, \quad \text{where} \quad \text{sim}( \mathbf{q}_i, \mathbf{k}_j) = \exp\left( \frac{\mathbf{q}^\top_i\mathbf{k}_j}{\sqrt{d}} \right) ,
     \label{eq1}
\end{equation}
where $ \text{sim}(\mathbf{q}_i, \mathbf{k}_j) $ measures the similarity between $i$-th query in $\mathbf{Q}$ and $j$-th key in  $\mathbf{K}$. Thus, the time and memory cost for the whole sequence is $\mathcal{O}(n^2)$, which limits its scalability to long sequences.
 
\subsection{The Proposed MGFormer}
In this section, we introduce our MGFormer for collaborative filtering.
Given a graph, we simply treat all nodes as independent tokens, augment them with positional encodings, and feed them into a masked kernel attention module. Next, we briefly present the process of MGFormer, including embedding lookup, positional encodings, masked kernel attention, and optimization.

\subsubsection{\textbf{Embedding Lookup}} The initial representations of a user $u$ and an
item $i$ can be obtained via embedding lookup tables:
\begin{equation}
 \begin{aligned}
	\mathbf{e}_{u}=\text{lookup}(u), \qquad \mathbf{e}_{i}=\text{lookup}(i),
 \label{eq2}
 \end{aligned}
\end{equation}
	where  $\mathbf{e}_{u} \in \mathbb{R}^d$ and $\mathbf{e}_{i} \in \mathbb{R}^d$ are the embeddings of user $u$ and item $i$, respectively, and $d$ is the embedding size. Then we concatenate all node  embeddings as: $\mathbf{E} \in \mathbb{R}^{(M+N) \times d}$.

\subsubsection{\textbf{Structural Encodings}} It is important to leverage the graph structural information into Transformer models~\cite{ying2021transformers,rampavsek2022recipe,hussain2022global,geisler2023transformers}. For recommendation, we simply design the structural encodings of users/items based on SVD  of the user-item interaction matrix:
\begin{equation}
 \begin{aligned}
	(\hat{\mathbf{U}} \sqrt{\mathbf{\Sigma}}) \cdot (\hat{\mathbf{V}}\sqrt{\mathbf{\Sigma}})^\top \leftarrow 
 \text{SVD} (\mathbf{R}),
 \label{eq333}
 \end{aligned}
\end{equation}
where we can regard $(\hat{\mathbf{U}} \sqrt{\mathbf{\Sigma}})  \in \mathbb{R}^{M \times d} $ as users' structural encodings, and $(\hat{\mathbf{V}} \sqrt{\mathbf{\Sigma}})  \in \mathbb{R}^{N \times d}$ 
as items' structural encodings. These structural encodings establish coordinate bases to preserve the global graph structures~\cite{dwivedi2021graph}. We then concatenate 
both users' and items' structural embeddings  as: $\mathbf{P} \in \mathbb{R}^{(M+N) \times d}$.  Finally, we can combine $\mathbf{E}$ and $\mathbf{P}$ as the input for the graph Transformer model:
\begin{equation}
 \begin{aligned}
	\mathbf{X} = [\mathbf{E} , \mathbf{P} ]\in \mathbb{R}^{(M+N) \times 2d}.
 \label{eq4}
 \end{aligned}
\end{equation}

\subsubsection{\textbf{Masked Kernel Attention}} Given  $\mathbf{X}$, we obtain the queries  with weight $\mathbf{W}_Q \in \mathbb{R}^{2d \times 2d}$, keys with $\mathbf{W}_K \in \mathbb{R}^{2d \times 2d}$, and values:
\begin{equation}
 \begin{aligned}
	\mathbf{Q} = \mathbf{X}\mathbf{W}_Q, ~~ \mathbf{K} = \mathbf{X}\mathbf{W}_K, ~~ \mathbf{V} = \mathbf{X},
 \label{eq5}
 \end{aligned}
\end{equation}
here we remove the feature transformation of  values for simplicity as suggested by~\cite{he2023simplifying}. Then,  we treat all nodes as independent
tokens and feed them into Transformers. However, the standard self attention in Eq. (\ref{eq1}) is not scalable for large graphs. 

To simplify the notation, we denote the input length $n=M+N$ and $m=2d$ in the rest of the paper.

\textbf{Linearized Attention:} Inspired by~\cite{choromanski2021rethinking,katharopoulos2020transformers}, we can use an arbitrary positive-definite kernel $\kappa(\cdot,\cdot)$ to serve as    $ \text{sim}(\cdot,\cdot) $, which can be further approximated by \textit{Random Features}, \textsl{i.e.}, $\kappa(\mathbf{a},\mathbf{b}) = \phi(\mathbf{a})^\top\phi(\mathbf{b})$, where $\phi(\cdot)$ is a  feature map. As such,   Eq. (\ref{eq1}) becomes:
\begin{equation}
\small
    \mathbf{h}_i = \frac{\sum_{j=1}^{n}\phi( \mathbf{q}_i)^\top \phi(\mathbf{k}_j) \cdot \mathbf{v}_j}{\sum_{j=1}^{n} \phi( \mathbf{q}_i)^\top \phi(\mathbf{k}_j)} = \frac{\phi( \mathbf{q}_i)^\top \cdot \sum_{j=1}^{n}\phi(\mathbf{k}_j)  \mathbf{v}^\top_j}{\phi( \mathbf{q}_i)^\top \sum_{j=1}^{n}  \phi(\mathbf{k}_j)}.
     \label{eq6}
\end{equation}

The crucial advantage of Eq. (\ref{eq6}) is that we can compute $\sum_{j=1}^{n}\phi(\mathbf{k}_j)  \mathbf{v}^\top_j$ and $\sum_{j=1}^{n}  \phi(\mathbf{k}_j)$ once and reuse them for every query, leading to a linear time and memory complexity.

\textbf{The choice of $\phi(\cdot)$:} The feature map $\phi(\cdot)$ can be some non-linear functions, such as $\textsl{elu}(\cdot)+1$~\cite{katharopoulos2020transformers}, $\textsl{relu}(\cdot)$ ~\cite{zhen2022cosformer}, or focused linear attention~\cite{han2023flatten}. Alternatively,  one can approximate the Softmax attention with \textsl{Random Fourier Features} (RFFs)~\cite{li2019towards},  \textsl{Orthogonal Random Features} (ORFs)~\cite{yu2016orthogonal} or \textsl{Positive Random Features} (PRFs)~\cite{choromanski2021rethinking}. Here we adopt recent proposed \textsl{Simplex Random Features} (SimRFs)~\cite{reid2023simplex} as the feature map. The  SimRFs $\phi(\cdot): \mathbb{R}^m \to  \mathbb{R}^m$ is defined  as:
\begin{equation}\label{ee6}
\phi(\mathbf{a})\overset{\mathrm{def}}{=}
\sqrt{\frac{1}{m}}\exp( \frac{-\|\mathbf{a}\|_2^{2}}{2} )[\exp(\mathbf{w}_{1}^{\top}\mathbf{a}), \cdots, \exp(\mathbf{w}_{m}^{\top}\mathbf{a})],
\end{equation}
where $\mathbf{W} = [\mathbf{w}_1, \cdots, \mathbf{w}_m] \in \mathbb{R}^{m \times m}$ is a random  matrix, which is:
\begin{equation*}\label{eqw}
\mathbf{W} = \mathbf{D} \mathbf{S} \mathbf{R}_o,
\end{equation*}
where $\mathbf{D}$ is a diagonal matrix with $\mathbf{D}_{ii}$ sampled from the  $\chi_m$-chi  distribution. $\mathbf{R}_o \in \mathbb{R}^{m \times m}$ is a random orthogonal matrix drawn from the Haar measure on $\mathrm{O}(m)$ \cite{yu2016orthogonal}. The rows $\boldsymbol{s}_i$ of the simplex projection matrix $\mathbf{S} \in \mathbb{R}^{m \times m}$ are given by:
\begin{equation*}
\label{eq:simplex_vectors}
\boldsymbol{s}_i = \begin{cases}
     \sqrt{\frac{m}{m-1}} \textbf{u}_i - \frac{\sqrt{m}+1}{(m-1)^{3/2}} (1,...,1,0)^{\top}  & \text{for}\ 1 \leq i < m, \\
      \frac{1}{\sqrt{m-1}}(1,1,...,1,0)^{\top} & \text{for}\ i = m, \\
    \end{cases}
\end{equation*}
where $\textbf{u}_i $ is the standard basis vector. The SimRFs has been demonstrated to yield the smallest mean square error in unbiased estimates of the Softmax kernel (refer ~\cite{reid2023simplex} for more details).

\textbf{Masked Attentions:} The graph topology is a strong inductive bias of graph data; beyond using it as the node positional encodings, it can be used to enhance the attention matrix, e.g., reweighing/masking the attention map with the adjacency/shortest-path matrix~\cite{huang2023tailoring,ying2021transformers}. Our kernelized attention Eq. (\ref{eq6}) can be also empowered  with a learnable topology-aware mask $\mathbf{M}$ to concentrate the distribution of attention scores. Here we introduce masked mechanism into Eq. (\ref{eq6}): 
\begin{equation}
\small
    \mathbf{h}_i = \frac{\sum_{j=1}^{n} \mathbf{M}_{ij} \phi( \mathbf{q}_i)^\top \phi(\mathbf{k}_j)  \cdot \mathbf{v}_j}{\sum_{j=1}^{n} \mathbf{M}_{ij}\phi( \mathbf{q}_i)^\top \phi(\mathbf{k}_j)} = \frac{ \phi( \mathbf{q}_i)^\top \cdot \sum_{j=1}^{n}\mathbf{M}_{ij}\phi(\mathbf{k}_j) \mathbf{v}^\top_j}{\phi( \mathbf{q}_i)^\top \sum_{j=1}^{n}\mathbf{M}_{ij}  \phi(\mathbf{k}_j)},
     \label{reatt}
\end{equation}
where $\mathbf{M} \in \mathbb{R}^{n \times n}$ is a mask\footnote{Note that achieving the reweighing mechanism does not necessarily require $\mathbf{M}$ to be a hard binarized matrix.} that would alter the attention distribution, assigning more credits to more crucial tokens.  Moreover, Eq. (\ref{eq6}) is a special case of Eq. (\ref{reatt}) by setting $\mathbf{M}$ as an all-ones matrix. 

 Unlike Eq. (\ref{eq6}), to obtain  outputs $\{\mathbf{h}_i\}^n_{i=1}$, we need to compute two matrices $\mathbf{P}_1 = \{\sum_{j=1}^{n}\mathbf{M}_{ij}\phi(\mathbf{k}_j) \mathbf{v}^\top_j\}_{i=1}^n$ and $\mathbf{P}_2 = \{\sum_{j=1}^{n}\mathbf{M}_{ij}\phi(\mathbf{k}_j) \}_{i=1}^n$, where $\mathbf{P}_1$ (similar for $\mathbf{P}_2$) can be decomposed as:
\begin{equation}
\small
  \begin{aligned}
        \mathbf{P}_1 & = \begin{pmatrix}
                \mathbf{M}_{11}  & \mathbf{M}_{12} & \cdots & \mathbf{M}_{1n} \\
                \mathbf{M}_{21}  & \mathbf{M}_{22}& \cdots & \mathbf{M}_{2n}\\
                \vdots  &\vdots &\ddots &\vdots & \\
                \mathbf{M}_{n1}  &\mathbf{M}_{n2} & \cdots &\mathbf{M}_{nn}
        \end{pmatrix}
        \begin{pmatrix}
                \vect(\phi(\mathbf{k}_1)  \mathbf{v}_1^\top)\\
                \vect(\phi(\mathbf{k}_2)  \mathbf{v}_2^\top)\\
                \vdots \\
                \vect(\phi(\mathbf{k}_n)  \mathbf{v}^\top_n)
            \end{pmatrix},\label{Eqn:D1} 
    \end{aligned}
    \end{equation}
where $\vect(\cdot)$ is the vectorization. Clearly, when the mask $\mathbf{M}$ is dense and requires explicit precomputation, calculating $\mathbf{P}_1$ and $\mathbf{P}_2$ results in a quadratic complexity, which  undermines the benefits of employing kernelized attention.  


In NLP domains, Relative Positional Encoding that encodes the distance between any two positions has achieved promising performance~\cite{raffel2020exploring,shaw2018self}.
In light of this analogy,  we propose a relative $\sin$-based degree centrality matrix as the mask, 
which relatively captures the node importance:
\begin{equation}\label{sine}
    \mathbf{M}_{ij} = \sin\left(\frac{\pi}{2} \times \frac{z_{\deg(i)} + z_{\deg(j)}}{2}\right),
\end{equation}
where $\deg(i)$ is the degree of node $i$, and $z_{\deg(i)} \in (0,1)$
is learnable node degree centrality, in which we use embedding lookup of $\deg(i)$, following a projection matrix with the sigmoid function to learn $z_{\deg(i)}$.
Within the range of $(0, \frac{\pi}{2})$, the monotonically increasing of sine function ensures that nodes with higher degree are considered to be more influential in the graph. Such information regarding node degree has proven to be  critical  in Graph Transformers~\cite{ma2023,ying2021transformers}.

Then, we can compute the module $\mathbf{M}_{ij} \phi( \mathbf{q}_i)^\top \phi(\mathbf{k}_j)$ in Eq. (\ref{reatt}) as:
\begin{equation}
\small
\begin{aligned}
   &   \mathbf{M}_{ij} \phi( \mathbf{q}_i)^\top \phi(\mathbf{k}_j) = \sin(\frac{\pi}{2} \times \frac{z_{\deg(i)} + z_{\deg(j)}}{2}) \phi( \mathbf{q}_i)^\top \phi(\mathbf{k}_j)  \\= & \left( \sin(\frac{\pi z_{\deg(i)}}{4} )  \cos(\frac{\pi z_{\deg(j)}}{4} ) + \cos(\frac{\pi z_{\deg(i)}}{4} )  \sin(\frac{\pi z_{\deg(j)}}{4} )  \right) \phi( \mathbf{q}_i)^\top \phi(\mathbf{k}_j) \\ = &\left(\phi( \mathbf{q}_i) \sin(\frac{\pi z_{\deg(i)}}{4} ) \right)^\top \left(\phi(\mathbf{k}_j)\cos(\frac{\pi z_{\deg(j)}}{4} )\right)+ \\ & \left(\phi( \mathbf{q}_i) \cos(\frac{\pi z_{\deg(i)}}{4} ) \right)^\top \left(\phi(\mathbf{k}_j)\sin(\frac{\pi z_{\deg(j)}}{4} )\right) \\ = & \phi^{\sin}( \mathbf{q}_i)^\top \phi^{\cos} (\mathbf{k}_j) + \phi^{\cos}( \mathbf{q}_i)^\top \phi^{\sin} (\mathbf{k}_j),
\end{aligned}
 \label{reatt3}
\end{equation}
where {$\phi^{\sin}( \mathbf{q}_i) = \phi( \mathbf{q}_i) \sin(\frac{\pi z_{\deg(i)}}{4} )$, $\phi^{\cos} (\mathbf{k}_j)=\phi(\mathbf{k}_j)\cos(\frac{\pi z_{\deg(j)}}{4} )$, $\phi^{\cos}( \mathbf{q}_i)=\phi( \mathbf{q}_i) \cos(\frac{\pi z_{\deg(i)}}{4} )$}, and {$\phi^{\sin} (\mathbf{k}_j)=\phi(\mathbf{k}_j)\sin(\frac{\pi z_{\deg(j)}}{4} )$}. In this way,  Eq. (\ref{reatt}) can be expressed as:

\begin{equation}
\small
\begin{aligned}
     \mathbf{h}_i =& \frac{\sum_{j=1}^{n} \mathbf{M}_{ij} \phi( \mathbf{q}_i)^\top \phi(\mathbf{k}_j)  \cdot \mathbf{v}_j}{\sum_{j=1}^{n} \mathbf{M}_{ij}\phi( \mathbf{q}_i)^\top \phi(\mathbf{k}_j)} \\
     =& \frac{\sum_{j=1}^{n} \phi^{\sin}( \mathbf{q}_i)^\top \phi^{\cos} (\mathbf{k}_j) \cdot \mathbf{v}_j + \sum_{j=1}^{n}\phi^{\cos}( \mathbf{q}_i)^\top \phi^{\sin} (\mathbf{k}_j)  \cdot \mathbf{v}_j}{\sum_{j=1}^{n} \phi^{\sin}( \mathbf{q}_i)^\top \phi^{\cos} (\mathbf{k}_j)  + \sum_{j=1}^{n}\phi^{\cos}( \mathbf{q}_i)^\top \phi^{\sin} (\mathbf{k}_j) }\\
     =& \frac{ \phi^{\sin}( \mathbf{q}_i)^\top \cdot \sum_{j=1}^{n} \phi^{\cos} (\mathbf{k}_j) \mathbf{v}^\top_j + \phi^{\cos}( \mathbf{q}_i)^\top \cdot \sum_{j=1}^{n} \phi^{\sin} (\mathbf{k}_j)   \mathbf{v}^\top_j}{\phi^{\sin}( \mathbf{q}_i)^\top\sum_{j=1}^{n}  \phi^{\cos} (\mathbf{k}_j)  + \phi^{\cos}( \mathbf{q}_i)^\top\sum_{j=1}^{n} \phi^{\sin} (\mathbf{k}_j) }.
\end{aligned}
     \label{reatt5}
\end{equation}

Clearly, our Eq. (\ref{reatt5}) bears a linear complexity   with respect to the input sequence length, similar to Eq. (\ref{eq6}). More importantly, it can implement an attention reweighing mechanism by leveraging the relative node degree information.

\subsubsection{\textbf{Optimization}} 

Recent studies~\cite{wang2020understanding,wang2022towards} identify two key properties highly related to the quality of embeddings: alignment and uniformity.  To achieve the two properties, we adopt DirectAU loss ~\cite{wang2022towards} to train the model parameters. That is:
\begin{equation}
\small
 \begin{gathered}
    \mathcal{L}_{\rm align} =\mathop{\mathbb{E}}_{(u, i)\sim p_{\rm pos}}\| \mathbf{h}_u- \mathbf{h}_i\|^2, \\
    \mathcal{L}_{\rm uniform} =\log\mathop{\mathbb{E}}_{(u,u')\sim p_{\rm user}}e^{-\|\mathbf{h}_u - \mathbf{h}_{u'}
        \|^2} + \log\mathop{\mathbb{E}}_{(i,i')\sim p_{\rm item}}e^{-\|\mathbf{h}_i  - \mathbf{h}_{i'} \|^2},\\
        \mathcal{L} = \mathcal{L}_{\rm align} + \lambda  \cdot \mathcal{L}_{\rm uniform},
    \label{eq12}
\end{gathered}
\end{equation}
where $\lambda$ is a regularized parameter and $\|\cdot\|$ indicates $l_2$ norm; $p_{\rm pos}$ denotes the distribution of positive user-item pairs; $ p_{\rm user}$ and $ p_{\rm item}$ are the distributions of users and items, respectively. Intuitively, 
linked user-item pair nodes should be close to each other
while random user/item nodes should scatter on the hypersphere.

Lastly, it is worth mentioning that our MGFormer guarantees the expressivity of learning all-pair interactions even using a single-layer single-head attention,  offering an advantage in capturing long-range dependencies compared to the GNN-based methods.

\begin{table}[]
\small
\caption{Statistics of three benchmark datasets.}
\label{tab:table1}
\begin{tabular}{lcccc}
    \toprule
Dataset  & \#user & \#item & \#inter. & density \\ \hline
Beauty   & 22.4k  & 12.1k  & 198.5k   & 0.07\%  \\
Yelp & 31.7k  & 38.0k  & 1561.4k  & 0.13\%  \\
Alibaba  & 106.0k & 53.6k  & 907.5k    & 0.016\% \\     \toprule
\end{tabular}
\vspace{-10pt}
\end{table}

\section{EXPERIMENTS}
\subsection{Experimental Settings}
\subsubsection*{{\textbf{Dataset}.}}  We conduct experiments on three  benchmarks~\cite{he2020lightgcn,huang2021mixgcf,wang2022towards}: \textbf{Amazon-Beauty}, \textbf{Yelp-2018}, and \textbf{Alibaba}.  The datasets are summarized in Table \ref{tab:table1}. We follow
the  strategy in~\cite{wang2022towards} to split the datasets into
training, validation, and testing sets.  To evaluate the performance, we adopt two common Top-$k$ metrics: Recall@$k$ and NDCG@$k$ ( $k=20$ by default) with  the all-ranking protocol~\cite{he2020lightgcn,wang2022towards}.

\subsubsection*{{\textbf{Baselines}.}}  We  choose the following baselines:  1) \textbf{ BPRMF}~\cite{rendle2009bpr}: a matrix factorization model. 2) \textbf{LightGCN}~\cite{he2020lightgcn}: a GNN model with linear propagation. 3) \textbf{SGL}~\cite{wu2021self}: a GNN model with contrastive learning. 4) \textbf{GOTNet}~\cite{chen2022graph}: a non-local GNN model. 5) \textbf{SimGCL}~\cite{yu2022graph}: a  contrastive model  with random augmentation. 6) \textbf{GFormer
}~\cite{Li_2023}:  a rationale-aware generative model. 7) \textbf{DirectAU}~\cite{wang2022towards}: a new loss to optimize alignment
and uniformity.

The embedding size $d$ is searched among $\{32, 64, 128\}$. The hyperparameters of all baselines are  carefully tuned to achieve the optimal performance. For DirectAU, we opt for LightGCN as the encoder. For 
MGFormer, a single-layer single-head attention is employed, as experiments indicate no substantial performance improvement with multi-layer multi-head attentions.  Also, we adjust the value of $\lambda$ in Eq. (\ref{eq12}) over the range from $0.1$ to $5.0$.

\begin{table}[]
\caption{The  performance for different models. The best results are in bold face, and the best baselines are underlined. }
\label{tab2}
\scalebox{0.80}{\begin{tabular}{l|cccccc}
\toprule[1.2pt]
         & \multicolumn{2}{c}{Beauty} & \multicolumn{2}{c}{Yelp}     & \multicolumn{2}{c}{Alibaba}       \\
Method   & recall          & ndcg            & recall          & ndcg            & recall          & ndcg            \\ \midrule
BPRMF~\cite{rendle2009bpr}    & 0.1153          & 0.0534          & 0.0693          & 0.0428          & 0.0439          & 0.0190          \\
LightGCN~\cite{he2020lightgcn} & 0.1201          & 0.0581          & 0.0833          & 0.0514          & 0.0585          & 0.0275          \\
SGL~\cite{wu2021self}      & 0.1228          & 0.0644          & 0.0896          & 0.0554          & 0.0602          & 0.0295          \\
GOTNet~\cite{chen2022graph}   & 0.1309          & 0.0650          & 0.0924          & 0.0567           & 0.0643          & 0.0301          \\
SimGCL~\cite{yu2022graph}   & 0.1367          & 0.0682          & 0.0937          & 0.0571          & {\ul 0.0667}    & {\ul 0.0311}    \\
GFormer~\cite{Li_2023}  & 0.1362          & 0.0671          & 0.0955          & 0.0597          & 0.0653          & 0.0305          \\
DirectAU~\cite{wang2022towards} & {\ul 0.1465}    & {\ul 0.0710}    & {\ul 0.0981}    & {\ul 0.0615}    & 0.0664          & 0.0308          \\ \midrule
MGFormer & \textbf{0.1531} & \textbf{0.0748} & \textbf{0.1051} & \textbf{0.0668} & \textbf{0.0702} & \textbf{0.0328} \\
Improv.  & +4.50\%         & +5.35\%         & +7.14\%         & +8.62\%         & +5.25\%         & +5.47\%         \\ \toprule[1.2pt]
\end{tabular}}
 \vspace{-15pt}
\end{table}

	\subsection{Experimental Results}
	\subsubsection*{\textbf{Overall Performance}}
	The results of different models in terms of Recall$@20$ and NDCG$@20$  are summarized in Table~\ref{tab2}. We find that our MGFormer generally outperforms the LightGCN and its variants by a large margin, with up to $8.62\%$ improvement.  This indicates that our one-layer kernelized attention model is indeed a powerful learner for link prediction in recommendation. 

For time complexity, both MGFormer and DirectAU theoretically achieve a linear complexity with respect to the number of nodes. As an example, DirectAU and MGFormer approximately take 3.6s and 8.7s per epoch for the Beauty dataset with the same hardware. The extra cost of MGFormer  is from computing the weights for queries and keys in Eq. (\ref{eq5}), and the random feature transformation in Eq. (\ref{ee6}). Overall, the results demonstrate the superiority of
our MGFormer. Specifically, it outperforms all baselines and maintains a comparable complexity for large-scale recommendation.

\begin{figure}
\centering
\includegraphics[width=6.6cm]{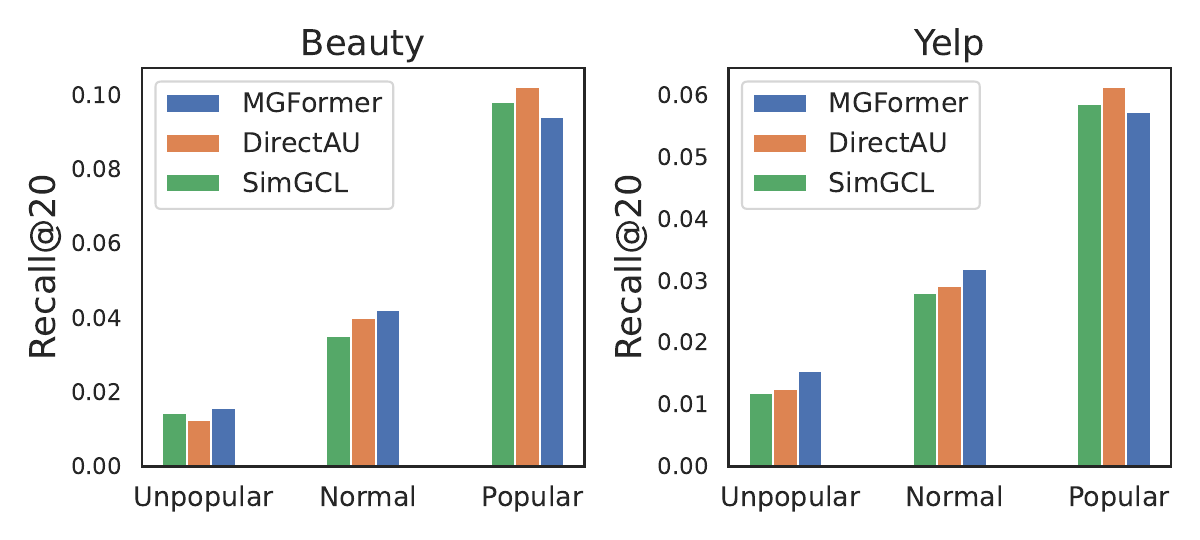}
\caption{Performance  for different item
groups.}
\label{srec}
\vspace{-15pt}
\end{figure}

	\subsubsection*{\textbf{Sparse Recommendation}} GNNs are known for their bias towards high-degree items, often neglecting the influence of low-degree items. Here we explore  the ability of capturing long-range dependencies on Beauty and Yelp datasets. Following~\cite{yu2022graph}, we divide the test set into three subsets based on the popularity of items: 'Unpopular', 'Normal', and 'Popular'. As shown in Figure~\ref{srec}, MGFormer outperforms GNN-based models for lower degree nodes, indicating the attention mechanism's proficiency in capturing long-range dependencies for sparse recommendation.

 	\subsubsection*{\textbf{Ablation Study}} We further explore several variants of MGFormer: 1) 
Remove structural encodings, 2) Remove sin-based degree centrality, 3) Choose graph adjacency as the mask~\cite{huang2023tailoring}, 4) Choose $1+elu(\cdot)$ as $\phi(\cdot)$~\cite{katharopoulos2020transformers}, 5) Choose focused function as  $\phi(\cdot)$~\cite{han2023flatten}. From Table~\ref{tab3}, we observe that: 1) The removal of structural encodings or degree centrality generally hampers performance. 2) The use of the adjacency matrix as a mask significantly diminishes performance, as the resulting attention becomes excessively sparse, thereby constraining its effectiveness. 3) Our model exhibits relative stability across various feature transformations.

\begin{table}[]
\caption{Ablation studies of MGFormer on Beauty dataset.}
\label{tab3}
\scalebox{0.86}{\begin{tabular}{l|cc}
 \toprule[1.2pt]
Version                                                                             & recall@20 & ndcg@20 \\ \midrule
default MGFormer                                                                            &    $0.1531$       &    $0.0748$     \\
w/o structural encodings                                                            &    $0.1413$       &      $0.0712$   \\
w/o degree centrality                                                               &   $0.1501$       &    $0.0739$     \\
choose graph adjacency as mask                                                            &    $0.1318$      &       $0.0674$  \\ 
choose $1+elu(\cdot)$ as $\phi(\cdot)$                                                     &      $0.1507$     &   $0.0740$      \\
\begin{tabular}[c]{@{}l@{}}choose focused function as  $\phi(\cdot)$\end{tabular} &     $0.1514$      &      $0.0742$   \\  \toprule[1.2pt]
\end{tabular}}
\vspace{-10pt}
\end{table}

\section{conclusion}
This study investigates the potential of a pure Transformer architecture in large-scale recommendation, where the scalability often poses a significant challenge. The core component, a masked kernelized attention module, allows us to attain a linear complexity. Interestingly, the experiments show that a single-layer attention model can deliver exceptionally competitive performance.

    \balance 
\bibliographystyle{ACM-Reference-Format}
\bibliography{sample-base}

\end{document}